\documentstyle[11pt,psfig]{article}
\def\be{\begin{equation}}
\def\ee{\end{equation}}
\begin{document}
\title{Renormalization group flow of SU(3) lattice gauge theory\\
- Numerical studies in a two coupling space -}

\author{QCD-TARO Collaboration\\
Ph.~de~Forcrand$^{\rm a}$,
M.~Garc{\'\i}a~P\'erez$^{\rm b}$,
T.~Hashimoto$^{\rm c}$,
S.~Hioki$^{\rm d}$,\\
H.~Matsufuru$^{\rm e}$,
O.~Miyamura$^{\rm f}$,
A.~Nakamura$^{\rm g}$,
I.-O.~Stamatescu$^{\rm h,i}$,\\
T. Takaishi$^{\rm j}$,
and
T.~Umeda$^{\rm f}$\\
\\
$^{\rm a}$Institute for Theoretical Physics, ETH-H\"onggerberg, CH-8093
Z\"urich, Switzerland \\
$^{\rm b}$Dept. F\'{\i}sica Te\'orica, Universidad Aut\'onoma de Madrid,
      E-28049 Madrid, Spain \\
$^{\rm c}$Dept. of Appl. Phys., Fac. of Engineering,
           Fukui Univ., Fukui 910-8507, Japan \\
$^{\rm d}$Dept. of Physics, Tezukayama Univ.,Nara 631-8501, Japan \\
$^{\rm e}$Research Center for Nuclear Physics, Osaka Univ.,
           Ibaraki 567-0047, Japan \\
$^{\rm f}$Dept. of Physics, Hiroshima Univ.,
           Higashi-Hiroshima 739-8526, Japan \\
$^{\rm g}$RIISE, Hiroshima Univ.,
           Higashi-Hiroshima  739-8521, Japan \\
$^{\rm h}$Institut f\"ur Theoretische Physik, Univ. Heidelberg
           D-69120 Heidelberg, Germany \\
$^{\rm i}$FEST, Schmeilweg 5, D-69118 Heidelberg, Germany \\
$^{\rm j}$Hiroshima University of Economics, Hiroshima 731-01, Japan}
%}

\date{}
\maketitle

\begin{abstract}

We investigate the renormalization group (RG) flow of SU(3) lattice gauge
theory in a two coupling space with couplings $\beta_{11}$ and
$\beta_{12}$
corresponding to $1\times 1$ and $1\times 2$ loops respectively.
Extensive numerical calculations of the RG flow are made in the fourth
quadrant
of this coupling space, i.e., $\beta_{11}>0$ and $\beta_{12}<0$.
Swendsen's factor two blocking and
the Schwinger-Dyson method are used to find an effective action for
the blocked gauge field.
The resulting renormalization group flow
runs quickly towards an attractive stream which has an approximate line
shape.
This is a numerical evidence of a renormalized trajectory
which locates close to the two coupling space.
A model flow equation which incorporates a marginal coupling
(asymptotic scaling term), an irrelevant coupling and a
non-perturbative attraction towards the strong coupling limit
reproduces qualitatively the observed features.
We further examine the scaling properties of an action which is
closer to the attractive stream than the currently used improved actions.
It is found that this action shows excellent restoration of rotational
symmetry even for coarse lattices with $a \sim 0.3$ fm.

\end{abstract}

\section{Introduction}
Since Wilson's first numerical renormalization group (RG) analysis of
SU(2) gauge theory \cite{Wilson},
there have been many Monte Carlo RG studies of non-perturbative
$\beta$-functions (see Ref.\cite{Gupta} and references therein).
In these analysis indirect information about the $\beta$ function,
such as $\Delta\beta$, has been obtained \cite{QCDTARO-MCRG}.
Recent progress of lattice techniques \cite{Okawa,HPW,Takaishi}
allows us to estimate directly the RG
flow in multi-coupling space \cite{patel_gupta}.

We study renormalization effects by means of a blocking transformation
which changes the lattice cut-off but leaves the long range contents of the
system invariant.
A new blocked action $S'$ as a function of blocked link variables $V$'s
is constructed from the original action $S(U)$ as
\be
e^{-S'(V)} = \int {\cal D}U e^{-S(U)} \delta(V-P(U))   ,
\ee
where $P$ defines the blocking transformation.
The action $S'$ includes the renormalization effects induced by
blocking.
In the space of coupling constants,
the blocking transformation makes a transition
from a point corresponding to $S$ to a new point, $S'$.
Repeating the blocking transformation, we obtain trajectories
in coupling space which define the so called renormalization group flow.

There is a special trajectory, i.e. renormalized trajectory (RT),
which starts at the ultra-violet fixed point. On the RT, the
long range information corresponding to continuum physics is preserved.
Recently Hasenfratz and Niedermayer have stressed that the action on the RT
can indeed be considered as a ``perfect action'' \cite{perfect}.
Therefore if we find a RT corresponding to a blocking transformation,
it provides an action which gives
accurate results corresponding to the continuum limit.
Even if it is an approximate one,
it serves as a well-improved action.  In this sense, a pioneering work
has been done by Iwasaki more than ten years ago \cite{Iwasaki0}.
He estimated a RT by matching Wilson loops based on a perturbative
approximation, and proposed an improved action which we will call below
Iwasaki action.

In this work, we analyze numerically the RG flow
in two coupling space, $(\beta_{11},\beta_{12})$,
of SU(3) lattice gauge theory and clarify the structure of
the renormalization group flow.
The action is restricted to the following form;
\begin{eqnarray}
S & = & \beta_{11} \sum_{{\rm plaq}}
(1 - {1 \over 3}{\rm Re \mbox{Tr}} P_{{\rm plaq}}) + \beta_{12} \sum_{{\rm
rect}}
            (1 - {1 \over 3}{\rm Re \mbox{Tr}} P_{{\rm rect}})  \nonumber \\
%  & = & - \frac{1}{2} a^4 ({\beta_{11}}+8 {\beta_{12}})
%  \sum {\rm \mbox{Tr}} F_{\mu\nu}(x)^2
%      + O(a^6) .
\label{eq:action}
\end{eqnarray}
Here $P_{{\rm plaq}}$ and $P_{{\rm rect}}$ correspond to
$1\times 1$ and $1\times 2$ loops, respectively.

The main purpose of the present work is
to perform an extensive study of the RT beyond a perturbative analysis.
It is a non-trivial fact whether the RT locates near the two coupling space,
i.e. in the $(\beta_{11}, \beta_{12})$ plane.
If no remnant of the RT can be seen in this plane, the two coupling space is
insufficient to obtain good improved actions.  In this sense,
a global analysis of the RG flow from weak to strong coupling regions
is indispensable.
%%%%%%%%%%%%%%%%%% Added  %%%%%%%%%%%%%%%%%%%%%%%%%%%%%%%%%%%%%%%%%%%%
Parts of our analysis were reported in Refs.\cite{QCDTARO96,Takaishi3},
and in this paper we present much more data which are enough to confirm
the existence of RT.
%%%%%%%%%%%%%%%%%%%%%%%%%%%%%%%%%%%%%%%%%%%%%%%%%%%%%%%%%%%%%%%%%%%%%%%

Our analysis is performed in the fourth quadrant of the coupling space.
We examine renormalization effects induced by Swendsen's factor two blocking
on the plaquette action as well as on some improved actions.  In addition,
we try to clarify the global structure of the RG flow.
An evidence that the RT sits close to the two coupling
space is provided by the fact that the flow runs quickly towards a narrow
attractive stream. Characteristic features of the flow in the
strong and weak coupling regions are also found.
The observed features are reproduced by
a model flow equation which incorporates a marginal coupling
(asymptotic scaling term), an irrelevant coupling and a
non-perturbative attraction towards the strong coupling limit.

Based on the flow structure, we examine the scaling properties of
several actions defined in this two coupling space.  Tests are made for
the rotational invariance and the scaling of $\sqrt{\sigma}/T_c$.
Near the attractive stream, we find good restoration of the rotational
invariance.

This paper is organized as follows.  In sect. 2, the basic tools for
the analysis are given. Sect. 3 is devoted to present
simulations and numerical results of the RG flow.  Scaling tests
are described in sect. 4.  In sect. 5, a model flow equation
which can reproduce the observed RG flow is proposed.
Renormalization effects beyond the two coupling space are discussed in
sect. 6.
 A summary of results is given in sect. 7.

\section{Blocking transformation and determination of renormalization
effects}

Here we describe the basic framework to study the RG flow
for SU(3) gauge fields on the lattice. First we
produce field configurations with an action $S$ which has coupling constants
$(\beta_{11},\beta_{12},\cdots)$, and apply a blocking transformation
on these configurations. Next we determine
an action $S'$ with coupling constants $(\beta '_{11},\beta '_{1
2},\cdots)$  which reproduces the transformed configurations. In this way,
we extract a flow in the coupling constant space
from the original coupling constants $\{\beta\}$ to the transformed ones
$\{\beta '\}$;
this is the RG flow.

%\subsection{Blocking procedure}
In order to perform the blocking transformation on a lattice,
we adopt Swendsen's factor-two blocking \cite{Swendsen}.
For a set of SU(3) link variables $U_{\mu}(n)$, a blocked field
is constructed as:
\begin{eqnarray}
&&Q_{\mu}(n) = U_{\mu}(n) U_{\mu}(n+\hat{\mu})
+ c \sum_{\nu \ne \mu} U_{\nu}(n) U_{\mu}(n+\hat{\nu})
                    U_{\mu}(n+\hat{\nu}+\hat{\mu})
U_{-\nu}(n+\hat{\nu}+2\hat{\mu}).
\end{eqnarray}
\noindent
Here $c$ is a parameter to control the weight of the staple-like paths.
A convenient notation $ V_{-\mu}(n) = V_{\mu}^{\dagger}(n-\hat{\mu}) $ is
also
used and the sum is taken over negative as well as positive direction.
\noindent
$Q_{\mu}(n)$  is projected onto the blocked SU(3) gauge field $V_{\mu}(n)$
by maximizing ${\rm Re} {\rm \mbox{Tr}}(Q_{\mu}(n) {V}_{\mu}^{\dag}(n))$.
\footnote{
The projection from $Q$ onto a SU(3) matrix $V$ is not unique.
In ref.~\cite{QCDTARO-MCRG}, we employed the polar decomposition,
i.e., $Q=V H$ where $H$ is an Hermite matrix and
$V=\tilde{V}/\mbox{det}(\tilde{V})$
with $\tilde{V}=Q/(Q^{\dagger}Q)^{1/2}$. One can easily prove that
both methods  are equivalent for unitary matrices $V$'s.
}

%\subsection{Schwinger-Dyson method to determine effective action}

To determine the effective action $S'$ on blocked
configuration $V$, we use
the Schwinger-Dyson method \cite{Okawa}.
This is based on the following identity:
for a link $V_{l_0}$, consider the quantities
\begin{eqnarray}
\langle  \mbox{Im} \mbox{Tr} (\lambda^b V_{l_0} G_{l_0}^{\alpha})\rangle
= \frac{1}{Z} \int {\cal D}V \mbox{ Im \mbox{Tr} }
(\lambda^b V_{l_0} G_{l_0}^{\alpha}) \
e^{-S'}
\label{SD0}
\end{eqnarray}
\noindent
where $\lambda^b$ stands for Gell-Mann matrices.
Here action is assumed to have the form,
\begin{eqnarray}
S' = \sum_{l}~ \sum_{\gamma} {\beta_{\gamma}' \over 2}
[1 - {1 \over 3}{\rm Re}~{\rm \mbox{Tr}} V_l G_l^{\gamma}]
\end{eqnarray}
and
$G_l^{\gamma}$ stands for a ``staple'' coupling to the link $l$
in a loop of type $\gamma$.
For the present analysis, $\gamma$ corresponds to a plaquette
and a rectangle.
Eq. (\ref{SD0}) should be invariant under the change of variable
$V_{l_0} \rightarrow (1+i\epsilon\lambda^{b})V_{l_0}$.
Setting terms linear
in $\epsilon$ to be zero, we get the identity,

\begin{eqnarray}
 &&\int {\cal D}V [
{\rm Re}{\rm \mbox{Tr}}((\lambda^b)^2 V_{l_0} G_{l_0}^{\alpha})
 + {\rm Im}{\rm \mbox{Tr}}(\lambda^b V_{l_0} G_{l_0}^\alpha ){\rm Im}
{\rm \mbox{Tr}}(\lambda^b V_{l_0} G_{l_0})]  \ e^{-S'} = 0  ,
\label{SD1}
\end{eqnarray}
where
\begin{eqnarray}
G_l = \sum_{\gamma} \frac{\beta_{\gamma}'}  {6} G_l^{\gamma}.
\end{eqnarray}

Summing over $b$ in the expression (\ref{SD1}) above, we obtain
the Schwinger-Dyson equation,

\begin{eqnarray}
&&{8 \over 3}{\rm Re}\langle {\rm \mbox{Tr}}(V_{l_0}
G_{l_0}^{\alpha})\rangle  =
 \sum_\gamma {\beta_{\gamma}' \over 6}
\{ -  {\rm Re} \langle {\rm
\mbox{Tr}}(V_{l_0}G_{l_0}^{\alpha}V_{l_0}G_{l_0}^{\gamma})\rangle
\nonumber
\\
&& +  {\rm Re} \langle {\rm
\mbox{Tr}}(G_{l_0}^{\alpha}(G_{l_0}^{\gamma})^{\dagger})\rangle
 -{2 \over 3} \langle {\rm Im~\mbox{Tr}}(V_{l_0}G_{l_0}^{\alpha})
{\rm Im~\mbox{Tr}}(V_{l_0}G_{l_0}^{\gamma})\rangle \}.
\label{eq:SD}
\end{eqnarray}
\noindent
Here we have used the identity: $
\sum_{b=1}^8 {\rm \mbox{Tr}}(\lambda^b A)
{\rm \mbox{Tr}}(\lambda^b B)=2{\rm \mbox{Tr}}AB-
\frac{2}{3}{\rm \mbox{Tr}}A{\rm \mbox{Tr}}B.
$
We apply eq. (\ref{eq:SD}) to the blocked configurations, and
calculate the expectation values $\langle \cdots\rangle $ on both sides. Now
eq. (\ref{eq:SD}) may be considered as a set of linear equations with
$\beta_\gamma$'s as unknowns.
It is noted that we may use other loop operators instead of $G^{\alpha}$.
However a minimal choice is to take the same $G^{\alpha}$s as the  ones
entering the
action. In this case,
the number of equations is equal to the number
of unknown couplings.

It is noted that the canonical Demon method also works for
the present purpose \cite{Takaishi}.  This method tunes the effective action
so as to reproduce
the mean values of the plaquette and rectangular loops whereas
the Schwinger-Dyson method respects wider loops which are combination of
staples
such as ${\rm \mbox{Tr}}(G_l^{\alpha}(G_l^{\gamma})^{\dagger})$.
In case of a limited coupling space, they may lead to different actions.
Because the systematic errors involved in both methods are not known in the
present stage,
only the results obtained via the Schwinger-Dyson method are presented in
this work.

Since we study the RG flow in the two coupling space, we show corresponding
Schwinger-Dyson equation explicitly.

\begin{eqnarray}
\left(
\begin{array}{l}
\beta'_{11} \\
\beta'_{12}
\end{array}
\right)=
\left(
\begin{array}{ll}
A_{11}&A_{12}\\
A_{21}&A_{22}
\end{array}
\right)^{-1}
\left(
\begin{array}{l}
\langle \mbox{Re Tr}(P^{(1)}_{\mu \nu})\rangle \\
\langle \mbox{Re Tr}(P^{(2)}_{\mu \nu})\rangle
\end{array}
\right)
\label{}
\end{eqnarray}
where
\begin{eqnarray}
\begin{array}{l}
A_{11}={1\over16}\sum_{\sigma\neq \mu}{\rm Re}
[
\langle \mbox{Tr}(P^{(1)}_{\mu \nu}P^{(1)\dagger}_{\mu \sigma})\rangle
-\langle \mbox{Tr}(P^{(1)} _{\mu \nu}P^{(1)} _{\mu \sigma})\rangle
-{1 \over 3}\langle \mbox{Tr}(P^{(1)} _{\mu \nu})
\mbox{Tr}(P^{(1)\dagger} _{\mu \sigma}-P^{(1)} _{\mu \sigma}
)\rangle ]
\\
A_{12}={1 \over 16}\sum_{\sigma \neq \mu}{\rm Re}
[
\langle \mbox{Tr}(P^{(1)} _{\mu \nu}P^{(2) \dagger} _{\mu \sigma})\rangle
-\langle \mbox{Tr}(P^{(1)} _{\mu \nu}P^{(2)} _{\mu \sigma})\rangle
-{1 \over 3}\langle \mbox{Tr}(P^{(1)} _{\mu \nu})
\mbox{Tr}(P^{(2)\dagger} _{\mu \sigma}-P^{(2)} _{\mu \sigma})
\rangle \\
\hspace{33mm}
+\langle \mbox{Tr}(P^{(1)} _{\mu \nu}P^{H \dagger} _{\mu \sigma})\rangle
-\langle \mbox{Tr}(P^{(1)} _{\mu \nu}P^{H} _{\mu \sigma})\rangle
-{1 \over 3}\langle \mbox{Tr}(P^{(1)} _{\mu \nu})
\mbox{Tr}(P^{H\dagger} _{\mu \sigma}-P^{H} _{\mu \sigma})
\rangle ]
\\
A_{21}={1\over16}\sum_{\sigma \neq \mu}{\rm Re}
[
\langle \mbox{Tr}(P^{(2)}_{\mu \nu}P^{(1) \dagger} _{\mu \sigma})\rangle
-\langle \mbox{Tr}(P^{(2)} _{\mu \nu}P^{(1)} _{\mu \sigma})\rangle
-{1 \over 3}\langle \mbox{Tr}(P^{(2)} _{\mu \nu})
\mbox{Tr}(P^{(1)\dagger} _{\mu \sigma}-P^{(1)} _{\mu \sigma})\rangle ]
\\
A_{22}={1 \over 16}\sum_{\sigma \neq \mu}{\rm Re}
[
\langle \mbox{Tr}(P^{(2)} _{\mu \nu}P^{(2)\dagger} _{\mu \sigma})\rangle
-\langle \mbox{Tr}(P^{(2)} _{\mu \nu}P^{(2)} _{\mu \sigma})\rangle
-{1 \over 3}\langle \mbox{Tr}(P^{(2)} _{\mu \nu})
{\rm Tr}(P^{(2)\dagger} _{\mu \sigma}-P^{(2)} _{\mu \sigma})
\rangle \\
\hspace{33mm}
+\langle \mbox{Tr}(P^{(2)} _{\mu \nu}P^{H\dagger} _{\mu \sigma})\rangle
-\langle \mbox{Tr}(P^{(2)} _{\mu \nu}P^{H} _{\mu \sigma})\rangle
-{1 \over 3}\langle \mbox{Tr}(P^{(2)} _{\mu \nu})
{\rm Tr}(P^{H\dagger} _{\mu \sigma}-P^{H} _{\mu \sigma})
\rangle ]
\end{array}
\end{eqnarray}

with
\begin{eqnarray}
\begin{array}{ll}
P^{(1)}_{\mu \nu}(n)=&
V_{\mu}(n)V_{\nu}(n+\mu)V_{-\mu}(n+\mu+\nu)V_{-\nu}(n+\nu)
\\
P^{(2)}_{\mu \nu}(n)=&V_{\mu}(n)
V_{\nu}(n+\mu)V_{\nu}(n+\mu+\nu)
V_{-\mu}(n+\mu+2\nu)V_{-\nu}(n+2\nu)
V_{-\nu}(n+\nu)
\\
P^{H}_{\mu \nu}(n)=&V_{\mu}(n)
[V_{\nu}(n+\mu)V_{-\mu}(n+\mu+\nu)
V_{-\mu}(n+\nu)V_{-\nu}(n-\mu+\nu)
V_{\mu}(n-\mu)
\\
&+V_{\mu}(n+\mu)V_{\nu}(n+2\mu)
V_{-\mu}(n+2\mu+\nu)V_{-\mu}(n+\mu+\nu)
V_{-\nu}(n+\nu)] \ \ .
\end{array}
\end{eqnarray}

\section{Simulation and numerical results of the RG flow}

Using the techniques in the previous section, we study the
coupling flow induced by the factor two blocking.
We set the blocking parameter $c=0.5$.

Series of simulations on lattices of size $8^4$ and $16^4$ are performed.
The region surveyed in the present work is
the fourth quadrant of the $(\beta_{11},\beta_{12})$ plane which covers
most improved actions presently known.
We use the pseudo heat-bath method to generate the gauge fields.
The blocking transformation is carried out at more than 30 points.
At each point, about 100 configurations separated by
every 100 sweeps are used to determine the renormalized couplings
$\beta_{11}'$ and $\beta_{12}'$ through the Schwinger-Dyson method.
Examples in the
determination of the coupling are shown
in Table \ref{tab:typical} and \ref{tab:deconf}.
The errors are given by the Jack-knife method and they are
relatively small even at deconfined points.
Analyses are made also for
data samples in which configurations are separated by 1000 sweeps.  Those
data agree with that of the standard samples within error bars.
An example is shown in the last two lines of Table \ref{tab:typical}.
\begin{table}
\begin{tabular}{|cc|cc|} \hline
$\beta _{11}$ & $\beta_{12}$ & $\beta _{11}'$ & $\beta_{12}'$  \\ \hline
7.00     & 0.0     & 13.188(22)  & -1.6564(94)  \\
7.00     & -0.35   & 8.148(29)   & -1.0431(98)  \\
11.00    & -0.9981 & 15.189(73)  & -2.329(19)   \\
13.20    & -1.493  & 15.445(50)  & -2.559(15)   \\
($\dag$)13.20    & -1.493  & 15.519(37)  & -2.583(13)  \\ \hline
\end{tabular}
\caption{Examples of renormalized couplings $\beta_{11}'$ and $\beta_{12}'$
obtained through the Schwinger-Dyson
method.  A hundred configurations are used at each point.
For ($\dag$) , measurements are performed every 1000 pseudo heat-bath steps,
while for the rest of the data we measure every 100 steps.}
\label{tab:typical}
\end{table}

We check also the effect of the finite temperature phase transition on the
determination of the coupling constants.
Since the blocking transformation induces renormalization effects
corresponding  to a change of lattice cutoff, the resulting coupling shifts
are
insensitive to the phases.
We examine this point
by comparing results belonging to the confined phase on the $16^4$ lattice
while remaining deconfined on the $8^4$ lattice at  several values of
$\beta$
as shown in Table \ref{tab:deconf}.
The finite temperature phase transition on the $8^4$ and $16^4$ lattices for
the
plaquette action takes place at approximately $\beta_{11}=5.9$ and $6.3$
which
roughly correspond
to $a\sqrt{\sigma}=0.27$ and $a\sqrt{\sigma}=0.14$ respectively.
As seen in Table \ref{tab:deconf}, both results agree within errors.

\begin{table}
\begin{tabular}{|cc|cc|c|} \hline
$\beta _{11}$ & $\beta_{12}$ & $\beta _{11}'$ & $\beta_{12}'$ &lattice size
\\ \hline
6.20     & 0.0     & 9.515(56)   & -1.156(23)  & $8^4$  deconfined \\
         &         & 9.547(14)   & -1.1671(53) & $16^4$ confined\\
\hline
9.8496     & -0.8937  & 12.368(61)  & -1.883(16) & $8^4$ deconfined \\
           &          & 12.280(14)   & -1.8688(41) & $16^4$ confined\\
\hline
9.0     & -1.03332  & 7.211(38)  & -1.046(17)  & $8^4$ deconfined \\
        &           & 7.2086(86) & -1.0456(30) & $16^4$ confined\\
\hline
\end{tabular}
\caption{Renormalized couplings $\beta'_{11}$ and $\beta'_{12}$
on the deconfined and confined lattices.
These are calculated with 50 configurations, and
measurements are performed every 100 steps.}
\label{tab:deconf}
\end{table}

\begin{figure}[hbt]
\center{
\leavevmode\psfig{file=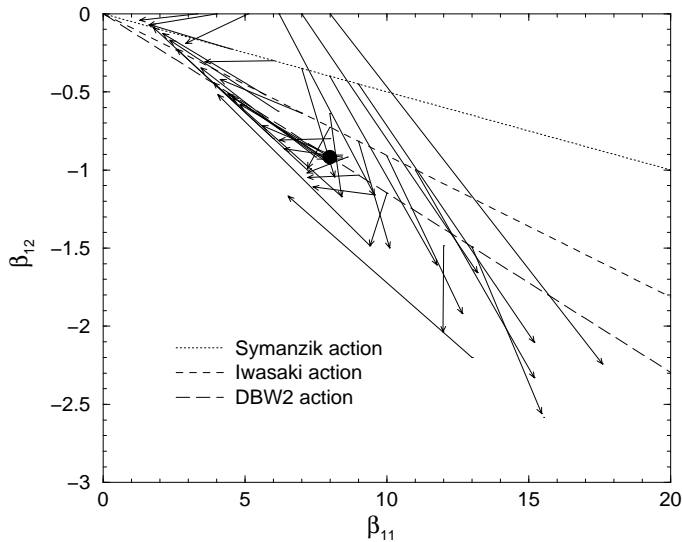,width=90mm,bblly=80bp}}
\caption{Renormalization group flow of SU(3) lattice gauge theory in
the two coupling space $(\beta_{11},\beta_{12})$.
There is an attractive stream  to which the arrows converge.
Dotted and dashed lines
correspond to ( tree level ) Symanzik
action and Iwasaki action respectively.
The long dashed line corresponds to the ``DBW2'' action
introduced in section 4. On the DBW2 line, the
black dot is the point reached by twice blocking
 from the plaquette action at $\beta=6.3$ on
a $32^3 \times 64$ lattice
\cite{Takaishi3}.
}
\label{fig:flow}
\end{figure}

Our results are summarized in Fig. \ref{fig:flow},
in which the coupling shift resulting after the blocking is indicated
by arrows.

%Fig1: RG Flow

As shown in the figure, several characteristics of the flow are seen.
If we start from the plaquette action ($\beta_{12}=0$ line),
renormalization results in a negative $\beta_{12}$ as expected by a
perturbative
analysis.   At $\beta_{11}=6 \sim 8$, the renormalization effect
is very strong making $\beta_{11}$ twice larger and $\beta_{12}$
negative.  The resulting points are far below the line of the tree-level
Symanzik action.
On the other hand, in the strong coupling region below $\beta_{11} <5$,
the effect of the renormalization is to reduce $\beta_{11}$.  Therefore, the
plaquette
action suffers large renormalization effects and it is far from the
renormalized
trajectory.

Points starting on the line defined by the tree-level Symanzik action,
$\beta_{12}/\beta_{11}=-0.05$,
are RG transformed onto ones with $\beta_{12}$ far more negative.
Although renormalization effects are reduced, the trend
is the same as that of the plaquette action.
This means that tree-level Symanzik action is still far from blocking
invariant.
This indicates that perturbative $O(a^2)$ improvement is insufficient
at least for the currently used range of lattice spacings.

Now we look at flows starting from points on the line corresponding
to Iwasaki action, i.e.,  $\beta_{12} = -0.09073 \beta_{11}$.
We see that renormalization occurs
approximately along the line up to an intermediate point
($\beta_{11} \approx 7.3 , \beta_{12} = -0.09073\beta_{11}$).
At larger $\beta_{11}$, however, flows depart from the line defined
by this action.
Thus the present blocking transformation renormalizes Iwasaki action
further and induces more negative values of $\beta_{12}$ above the
intermediate
coupling region.

Let us turn to the global structure of the flow.
Blockings are made starting from  points  with
$\beta_{12}/\beta_{11}= -0.1 \sim -0.15$.  For those points, the
renormalization
effect is relatively small and converges to a narrow stream.  This
 trend is manifest at strong coupling.
As a whole, there is an attractive stream which
the flow approaches quickly.  Furthermore, once the flow reaches the
stream, it
runs along it. Therefore actions on the attractive stream are
approximately blocking invariant apart from a normalization.
The shape of the attractive stream is clearly recognized
as a parabolic curve in the strong coupling region while
at points far from the origin, it is less obvious with the present
data.
This is a remarkable indication that
the renormalized trajectory locates close to
the $(\beta_{11},\beta_{12})$ coupling space.
This is an encouraging result for finding a good improved action
in this two coupling space.

A closer look at the attractive flow allows to extract more information.
If we start from the Wilson action at $\beta_{11}>6.0$, the first blocking
leads to larger renormalization effects as $\beta_{11}$ increases and
the resulting flow vectors have the same direction pointing towards
increasing
$\beta_{11}$ further until the flow has reached the main attractive stream.
This feature seems consistent with a flow induced by an irrelevant coupling.
In the strong coupling region the behavior is quite different,
under the blocking the coupling moves deeper into strong coupling, with the
main stream
following a parabolic behavior, as already indicated above.
In section 5, we will try to reproduce these features by a model equation
including
an irrelevant coupling, an asymptotic scaling term and
a driving term derived from the area law behavior of Wilson loops.

\section{Tests of improvement for actions near the attractive stream}

The purpose of this section is to examine the degree of improvement for
actions lying near the attractive stream.  As shown in Fig. \ref{fig:flow},
we expect that the irrelevant coupling is considerably reduced
and actions will be dominated by the marginal coupling term.
Here we
analyze an action suggested by double blocking from
Wilson action at $\beta=6.3$ on a $32^3 \times 64$ lattice
where ($\beta_{11},\beta_{12})=(7.986(13),-0.09169(41))$
has been obtained \cite{Takaishi3}.
This action is called DBW2 (doubly blocked from Wilson action in
two coupling space ) and
has a value of $\beta_{12}/\beta_{11}$  equal to -0.1148
(long dashed line in Fig. \ref{fig:flow}).

There are many ways to test improvement.  Among them, we
will examine rotational invariance of the heavy quark potential and
independence of $T_c/\sqrt{\sigma}$ on the lattice spacing.
Let us define a measure of violation of rotational symmetry
as

\be
\delta_V^2=\sum_{{\rm off}}{[V(R)-V_{{\rm on}}(R)]^2 \over V(R)^2 \delta
V(R)^2}
/(\sum_{{\rm off}}{1 \over \delta V(R)^2})
\label{rotvio}
\ee
\noindent
where $V(R)$ is the static quark potential and $\delta V(R)$ is its error.
$V_{\rm on}(R)$ is a fitting function to only on-axis data.  $\sum_{{\rm
off}}$
means summations over off-axis data.  This quantity is measured
on the configurations generated by the DBW2 action.
For comparison, plaquette-, tree-level Symanzik and Iwasaki improved
actions are also examined.  A $12^3 \times 24$ lattice is used for this
purpose
and simulations are made for lattice spacings ranging from  $a=0.15$ to $
\sim 0.4$fm.
The statistics is a hundred configurations for each data point and
the error is given by the Jack-knife method.

\begin{figure}[hbt]
\center{
\leavevmode\psfig{file=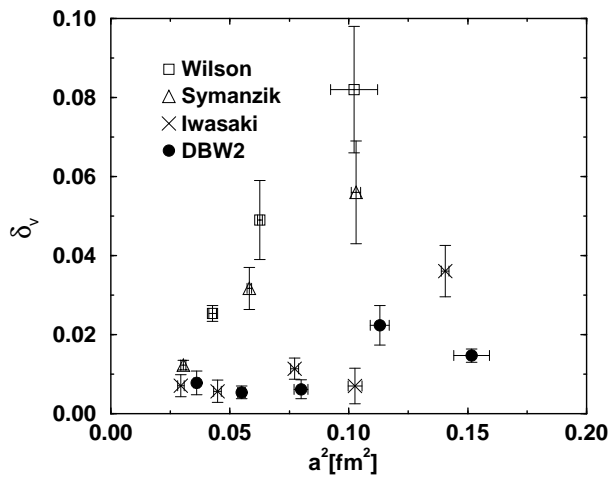,width=80mm,bblly=80bp}}
\caption{Rotational symmetry violation $\delta_V$ versus $a^2$ for
various improved actions.}
\label{fig:rot}
\end{figure}

The results are summarized in Fig. \ref{fig:rot}.

%Fig. 2 Rotational symmetry

In the figure, the horizontal axis is the
lattice spacing squared so as to see the expected $O(a^2)$ violation.
  As seen in
the figure, both Iwasaki  and DBW2 actions show excellent
restoration of rotational symmetry even at $a \sim 0.3$fm while
clear $a^2$ violations are seen for plaquette and tree-level Symanzik
actions.

A second check is the scaling of $T_c/\sqrt{\sigma}$.
The critical temperature is defined as
\be
T_c=1/N_t a_c \ , \ \ \ a_c=a(\beta_c)
\label{tc}
\ee
where $N_t$ is the temporal lattice size.
In order to obtain the critical
coupling $\beta_c$ , $N_t=3,4,6$ lattices are used and the Polyakov loop
susceptibility is measured.  Here, histogram method is utilized
to determine the peak of the susceptibility.\cite{Histogram}
Then, $\beta_c$ at infinite volume limit is obtained by
finite size scaling of $12^3 \times 4$ and $16^3\times 4$ lattices.
The resulting $\beta_c$ are given in Table \ref{tab:beta_c}.

\begin{table}
\begin{tabular}{|c|c|c|} \hline
$N_t$ & $N_s$  & $(\beta_{11}+8\beta_{12})_c \ \ \mbox{at} \ \ V=\infty$
\\ \hline
3     & 10, 12, 14 & 0.75696(98)  \\
4     & 12, 16     & 0.82430(95)  \\
6     & 18         & 0.9636(25)    \\ \hline
\end{tabular}
\caption{Phase transition points for the DBW2 action at
finite temperature. Results are obtained by extrapolating
to the infinite volume limit.}
\label{tab:beta_c}
\end{table}

The string tension is measured on $12^3\times 24$ and $18^3\times 36$
lattices at
the values of the coupling determined by the $T_c$ analysis. It
is extracted
from the static quark potential using the Ansatz ;
\be
V(R)= A +{\alpha \over R} + \sigma R
\label{pot}
\ee

\begin{table}
\begin{tabular}{|c|c|c|c|} \hline
$\beta_{11}+8\beta_{12}$ & $A$      & $\alpha$  & $\sigma a^2$   \\ \hline
  0.82430   &0.550(17) &-0.255(23) & 0.1555(28)     \\
  0.9636   &0.5791(96) &-0.357(22) & 0.06996(99)    \\ \hline
\end{tabular}
\caption{Parameters of the heavy quark potential for the DBW2 action.
These simulations are carried out on
$12^3\times 24$ and $18^3\times 36$  lattices while
the values of the coupling are those determined by $T_c$ analysis (Table
\ref{tab:beta_c}). Statistics is 130 configurations per data at
$\beta_{11}+8\beta_{12}=0.82430$
and 80 configurations per data at $\beta_{11}+8\beta_{12}=0.9636$.
}
\end{table}

\begin{figure}[hbt]
\center{
\leavevmode\psfig{file=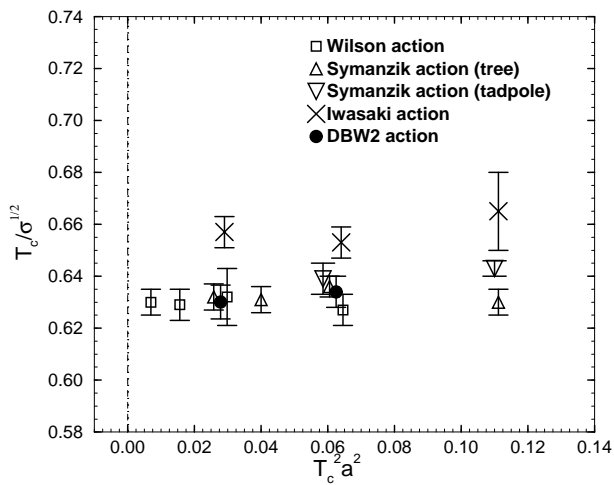,width=80mm}}
\caption{Scaling behavior of $T_c/\sqrt{\sigma}$. We compare the DBW2
action
with results from other actions in two-coupling space \cite{kaneko,string}.}
\label{fig:scale}
\end{figure}

From the extracted values of $\sigma$ at $\beta_c(N_t=3)$, $\beta_c(N_t=4)$
and $\beta_c(N_t=6)$
we obtain the following values of $T_c/\sqrt{\sigma}$:
\begin{eqnarray}
T_c/\sqrt{\sigma}=& 0.6340(60)\ \ \ {\rm at}\ \ \ N_t=4 \nonumber \\
                  & 0.6301(65)\ \ \ {\rm at}\ \ \ N_t=6
\end{eqnarray}
\noindent
In Fig. \ref{fig:scale} we show the obtained results together with
other data from plaquette (Wilson), tree-level Symanzik,
tadpole improved Symanzik and Iwasaki actions taken from Refs
\cite{kaneko,string}.

%Fig. 3 $T_c/\sqrt{\sigma}$ versus $a^2$

No appreciable dependence on the lattice spacing is seen in the ratio
$T_c/\sqrt{\sigma}$ for
all the cases including the DBW2 action.  
%A puzzle is that the value
%for the Iwasaki action is slightly higher in comparison with other data but
%this may be due to technical differences in the way the string tension is
%extracted.

The results of both tests is an indication that the DBW2 action is
a well-improved action as suggested by the
non-perturbative RG flow analysis.

\begin{figure}[hbt]
\center{
\leavevmode\psfig{file=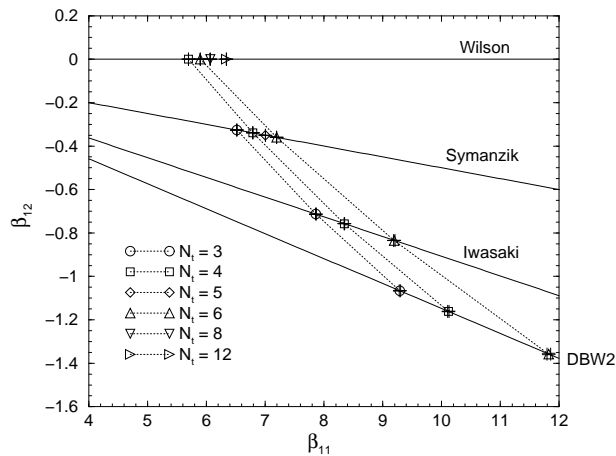,width=80mm}}
\caption{ Contours of constant $aT_c$ in the
($\beta_{11},\beta_{12}$)-plane.
Symbols represent the phase transition points for each $N_t$
at the thermodynamic limit.
Dotted lines connect the points with the same $aT_c (=1/N_t)$ and
they suggest contours of $aT_c=const.$.}
\label{fig:tc}
\end{figure}

As a by-product of the $T_c$ analysis, contours of constant
lattice spacing are obtained through the relation $a=1/N_t T_c$.
Fig. \ref{fig:tc} shows a compilation of the phase transition points in the
two coupling space for the plaquette,
tree-level Symanzik, Iwasaki and DBW2 actions on $N_t=3 \sim 12$ lattices.

%Fig. 4  $aT_c =$ constant

%%%%%%%%%%%%%%%  this section has been shorten %%%%%%%%%%%
\section{A model analysis for the flow}
Results of the RG flow in the section 3 provide basic
information on relation between couplings and lattice spacing.
In order to understand the driving force of the flow,
we examine a toy model which
incorporates a perturbative scaling term
and an irrelevant coupling
at
small lattice spacing and an attraction originated from area law
in strong coupling region.

We consider two dimensional model beta function which has a marginal
coupling and an irrelevant one at small lattice spacing as
\be
{d \vec{\beta} \over d \ln a} = {\bf A} \vec{\beta} - B(\beta_n)
{\vec{n} \over |\vec{n}|^2}
\label{weq1}
\ee
where $B(x)$ is the perturbative beta-function
\be
B(x)=12 b_0 + 72 \frac{b_1}{x} + \cdots
\label{weq2}
\ee
with $b_0=33/(48\pi^2)$ and $b_1=(102/121) b_0^2$.
${\bf A}$ is a constant $2\times 2$ matrix which has
a zero eigenvalue with an eigenvector $\vec{w}$ and a
negative eigenvalues
$\lambda$ with a eigenvector $\vec{v}$ .
$\vec{n}$ is constant vector orthogonal to $\vec{v}$ , i.e.,
$(\vec{n}\cdot \vec{v})=0$ .
By this condition, resultant solution exibits correct scaling
property for $(\vec{n}\cdot \vec{\beta})$.

Solution of eq. (\ref{weq1}) is
\be
\vec{\beta} =( c a^{\lambda}+ d(a))
\vec{v} + {\beta_p(a\Lambda) \over (\vec{n}\cdot \vec{w})}\vec{w}
\label{ansol}
\ee
where $\beta_p(a\Lambda)$ is the asymptotic scaling solution satisfying
\be
{d\beta_p \over d \ln a}=-B(\beta_p)
\ee
\noindent
and
\be
d(a)= {(\vec{v}\cdot \vec{w}) \over
(\vec{n}\cdot \vec{w})} \int^1_0
d\xi \ \xi^{-\lambda-1} B(\beta_p(\xi a\Lambda)) \quad .
\ee
It is noted that $(\vec{n}\cdot \vec{\beta})= \beta_p(a\Lambda) .
$

Attractive driving force in the strong coupling region
comes in from the string tension.
The lowest order of
strong coupling calculation for $N\times M$ Wilson loops is
(suppose that $NM$ is an even integer)
\begin{eqnarray}
< W(N\times M) > = (\beta_{11} /18 )^{NM} +
(\beta_{11} /18 )^{NM-2}(\beta_{12} /18 ) P^{NM}_1 + \nonumber \\
(\beta_{11} /18 )^{NM-4}(\beta_{12} /18 )^2 P^{NM}_2 + \cdot \cdot
+(\beta_{12} /18 )^{NM/2} P^{NM}_{NM/2}
\label{st0}
\end{eqnarray}
\noindent
where $P^{NM}_k$ are the tiling weights for filling the area of the loop by
$(NM-2k) \ [1\times 1]$ and $k \ [1\times 2]$ tiles.
Then, area law for the Wilson loop leads
\be
\beta_{11} /18  = \exp( -a^2 \sigma) , \ \ \ \
\beta_{12} /18  = C \exp( -2 a^2 \sigma) \ \ .
\ee
\noindent
In a differential form, we have
\begin{eqnarray}
{d \vec{\beta} \over da^2} = - \sigma
\left( \begin{array}{cc}
 1& 0 \\
 0 & 2\\
\end{array}
\right)
\vec{\beta} .
\label{st2}
\end{eqnarray}

Driving force in a two dimensional toy model beta function is assumed as
sum of those in the weak and strong-coupling region,i.e.,
\begin{eqnarray}
{d \vec{\beta} \over ds} = - 2s
\left( \begin{array}{cc}
 1+{\zeta_1 \over s} & 0 \\
 0 & 2+{\zeta_2 \over s} \\
\end{array}
\right)
\vec{\beta} +
{1 \over s }
[{\bf A} \vec{\beta} - B(\beta_n){
\vec{n} \over |\vec{n}|^2}] \ \ ,\ \ \ \ s \equiv a\sqrt{\sigma} \ \ .
\label{model1}
\end{eqnarray}
\noindent
Here, we keep the next order in $a$ in the non-perturbative term
as free parameters, $\zeta_1$ and $\zeta_2$.
Parameters included are $\theta , \theta' ,
\lambda , \zeta_1$ and $\zeta_2$.
$\vec{v}$ and $\vec{w}$
are specified
as
$\vec{v}=(\cos\theta',\ \sin\theta')$ and $\vec{w}=(\cos\theta,\
\sin\theta)$.
Here the matrix ${\bf A}$ is
${\bf A} = \lambda \vec{v} \otimes \vec{w_T}^t
/(\vec{v}\cdot\vec{w_T})
$ with $\vec{w_T}=i\sigma_2 \vec{w}$.
$\vec{n}$ is given by $\vec{n}=(1,\cot(-\theta'))$.

\begin{figure}[hbt]
\center{
\leavevmode\psfig{file=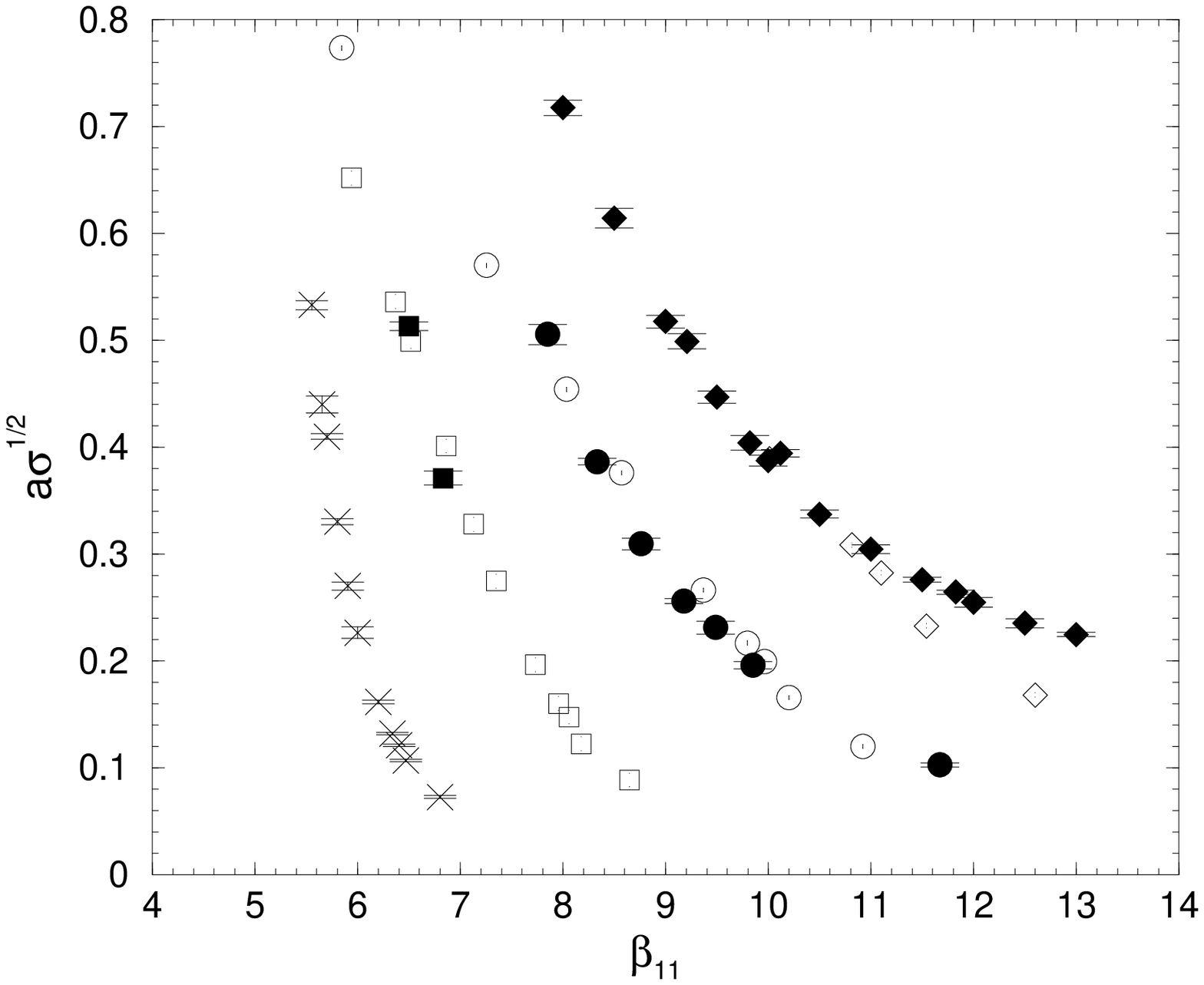,width=60mm}
\psfig{file=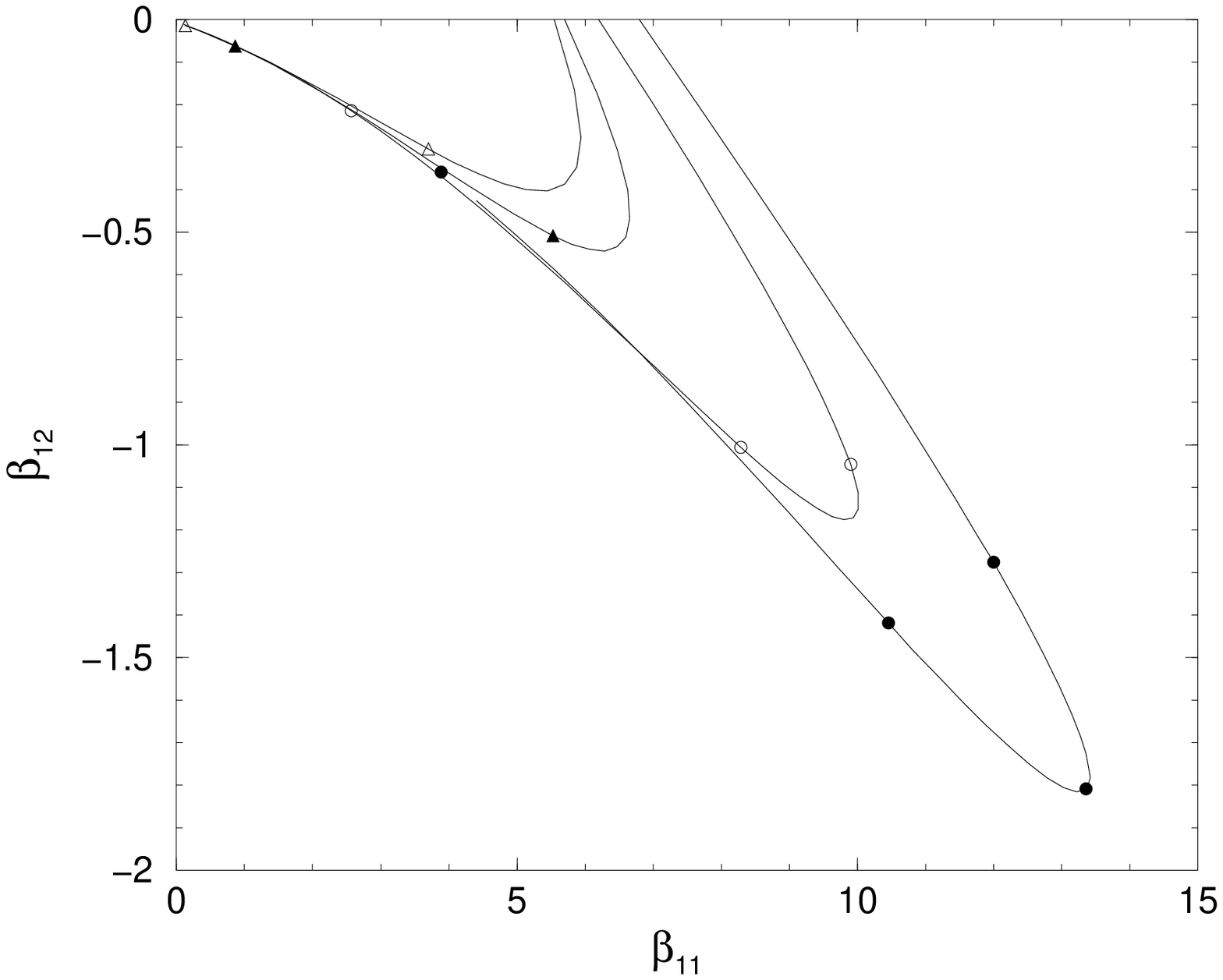,width=60mm}}
\caption{(a) Lattice spacing versus $\beta_{11}$.
Crosses are input data from
the plaquette action
given by \cite{Bali,Boyd,CPPACS1}.
Filled symbols are the data of tree-level Symanzik action (squares)
and DBW2 action (diamonds) \cite{Borici}. The corresponding open symbols
are the results predicted by eq. (\ref{model1}).
(b) Flow trajectories calculated by eq. (\ref{model1}).
The trajectories start at
 $(\beta_{11},\beta_{12}) = (5.55,0.0), (5.70,0.0), (6.20,0.0)$
 and $(6.80,0.0)$.
Symbols on the trajectories indicate the points corresponding to the
factor two blocking where the lattice spacing is
$2^n a_0 ( n=1,2,3,\cdots )$ with initial value $a_0$.
}
\label{fig:avsb}
\end{figure}

\begin{figure}[hbt]
\center{
\leavevmode\psfig{file=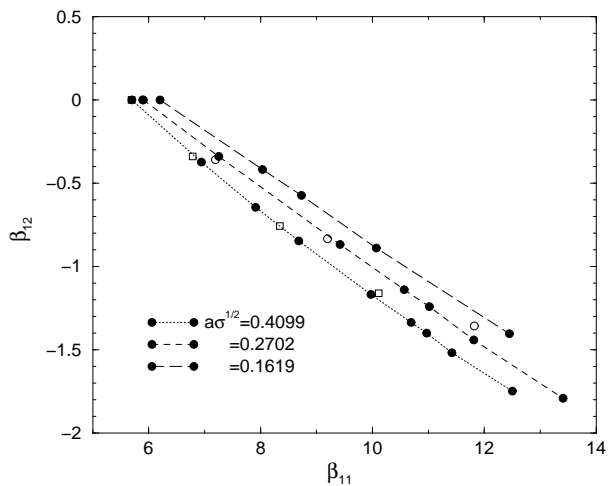,width=80mm}}
\caption{Contours of constant lattice spacing.
Filled circles show
the points with $a\sqrt{\sigma}=0.4099$ , $0.2702$ and $0.1619$
as computed by eq. (\ref{model1}).
% with initial conditions as in Table \ref{tab:init}.
The data for $(aT_c)^{-1}= 4$ (open squares) and 6 (open circles)
obtained by the scaling analysis (Fig. \ref{fig:tc})
are also indicated.}
\label{fig:cont}
\end{figure}

By this model, we calculate lattice spacings for different actions,
RG flow trajectories and contours of constant lattice spacings.
Comparisons with the present data and those reported in references
\cite{CPPACS,Borici} are made.
Giving initial conditions for $\beta_{11},\beta_{12}$ and $a$
on the $\beta_{12}=0$ line (plaquette action)
\cite{Bali,CPPACS1,Boyd} , those quantities
are calculated and shown in Fig.s \ref{fig:avsb},
 and \ref{fig:cont}.
%%%%%%%%%%%%%%%%%%%%%%%
\footnote{
$\sqrt{\sigma}=$420MeV is assumed for the data in reference
\cite{CPPACS1}.)}
%%%%%%%%%%%%%%%%%%%%%%%
Parameters are chosen so as to match the recently reported data
of $a\sqrt{\sigma}$ for the Iwasaki action \cite{CPPACS}
\be
\theta = -0.156 \ ,\ \ \theta' = -0.205 \ , \ \ \lambda=-0.5 \ ,
\ \ \zeta_1=0.1 \ \ \zeta_2=0.0 \ .
\label{5param}
\ee

As shown in the figures, lattice spacings in the $\beta_{11}-
\beta_{12}$ plane and the RG flow trajectories are reasonably
well reproduced.  As for the flow,
rapid approach to the attractive stream
is driven by the irrelevant coupling.
At intermediate lattice spacings, both the non-perturbative  and
asymptotic scaling terms drive the trajectories.  Finally, in the strong
coupling region, the non-perturbative term dominates and the parabolic
behavior of the trajectories sets on.
We also note that the model describes contours of constant
lattice spacing fairly well as seen in Fig.\ref{fig:cont}.

Through the study,
an understanding for basic driving mechanism of the RG flow
in $\beta_{11}-\beta{12}$ plane
by scaling term and an irrelevant coupling at
small lattice spacing and an attraction originated from area law
in strong coupling region is suggested.

%%%%%%%%%%%%%%%%%%%%%%%%%%%%%%%%%%%%%%%%%%%%%%%%%%%%
%Fig. 8  Lattice spacing versus $\beta_{11}$.
%%%%%%%%%%%%%%%%%%%%%%%%%%%%%%%%%%%%%%%%%%%%%%%%%%%%
%%%%%%%%%%%%%%%%%%%%%%%%%%%%%%%%%%%%%%%%%%%%%%%%%%%%
%Fig. 6  Flow trajectories described by the model
%\begin{figure}[hbt]
%\center{
%\leavevmode\psfig{file=fig6.eps,width=80mm}}
%\caption{Flow trajectories calculated by eq. (\ref{model1}).
%The trajectories start at
% $(\beta_{11},\beta_{12}) = (5.55,0.0), (5.70,0.0), (6.20,0.0)$
% and $(6.80,0.0)$.
%Initial values of $a\sqrt{\sigma}$ are given by those listed
%in Table \ref{tab:init}.
%Symbols on the trajectories indicate the points corresponding to the
%factor two blocking where the lattice spacing is
%$2^n a_0 ( n=1,2,3,\cdots )$ with initial value $a_0$.}
%\label{fig:flow_mod2}
%\end{figure}
%%%%%%%%%%%%%%%%%%%%%%%%%%%%%%%%%%%%%%%%%%%%%%%%%%%%
%%%%%%%%%%%%%%%%%%%%%%%%%%%%%%%%%%%%%%%%%%%%%%
%Fig. 7 Contours of constant lattice spacing.
%%%%%%%%%%%%%%%%%%%%%%%%%%%%%%%%%%%%%%%%%%%%%%

\section{Examination of renormalization effects outside of the two
coupling space}

In this work, the working space is limited to two coupling space
$(\beta_{11},\beta_{12})$ .  We have seen an evidence that
the RT locates close to that plane. Since this is a highly non trivial fact,
the truncation effect which comes from the limitation of the coupling space
to the two-dimensional plane should be examined.
There are several indirect indications that
truncation effects may be important. As seen in Fig. \ref{fig:flow},
the shape of the
attractor is less clear above $\beta_{11}\sim 12$ .  This suggests that
the distance to the RT grows in this region.

%Another point is that the lattice spacing of lattice which is generated
%by truncated couplings $$ after blocking
%is larger (roughly 40-60\%) than that of the original blocked lattice.

Another point is that the resulting couplings
$(\beta_{11}',\beta_{12}')$ obtained through the truncation
into two coupling space
%Schwinger-Dyson equation
after blocking
give larger lattice spacing (roughly 40-60 \%) than
that of the original blocked lattice.
%twice the original lattice spacing.
One possible reason is
the choice of a minimal set of loops in the Schwinger-Dyson equation.
But this effect might also be due to the truncation of the
space of couplings. See ref. \cite{Takaishi2}.

\begin{figure}[hbt]
\center{
\leavevmode\psfig{file=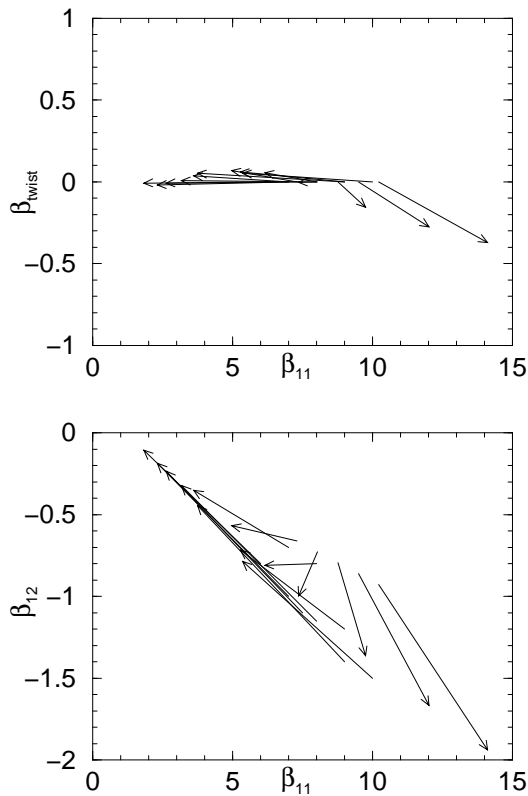,width=70mm}}
\caption{RG flow of SU(3) lattice gauge theory in three coupling
space $(\beta_{11},\beta_{12},\beta_{\rm twist})$.
Blocking is applied for the points near the attractive stream
on the $(\beta_{11},\beta_{12})$-plane at $\beta_{\rm twist}
=0$. The top figure shows the projection of the flow
onto the $(\beta_{11},\beta_{\rm twist})$
-plane while
the bottom figure is the projection onto the $(\beta_{11},\beta_{12})$
-plane.
}
\label{fig:3dim}
\end{figure}

In order to examine renormalization effects outside to the
two coupling space, we perform a limited analysis
in a three coupling space, $(\beta_{11},\beta_{12},\beta_{\rm twist})$,
with a twisted loop included in the action \cite{TARO2},

\begin{eqnarray}
S_{\rm twist} = &\beta_{\rm twist} \sum_{n, \mu \neq \nu \neq \tau}
[ 1 - {1 \over 3}  {\rm Re} \mbox{\rm Tr}( P_{\rm twist}^{\mu \nu \tau} )]
\end{eqnarray}
\noindent
where
\be
P_{\rm twist}^{\mu \nu \tau}=
U_{\mu}(n)U_{\nu}(n+\hat{\mu})U_{\tau}(n+\hat{\mu}+\hat{\nu})
U_{-\nu}(n+\hat{\mu}+\hat{\nu}+\hat{\tau})
U_{-\mu}(n+\hat{\mu}+\hat{\tau})
U_{-\tau}(n+\hat{\tau}) .
\ee
%%%%%%%%%%%%%%%%%% Added  %%%%%%%%%%%%%%%%%%%%%%%%%%%%%%%%%%%%%%%%%%%%
\noindent
We will study $\beta_{\rm twist}$ in addtion to $(\beta_{11},\beta_{12})$
because in Ref.\cite{Takaishi3} more general actions were studied and it
was found that the value of the twist coupling is larger that that of
the chair type loop.  Moreover the action with $W_{1\times 1}$,
$W_{1\times 2}$ and $W_{\rm twist}$ is often used in the reteratures.
\cite{Gupta,Lapage}.
%%%%%%%%%%%%%%%%%%%%%%%%%%%%%%%%%%%%%%%%%%%%%%%%%%%%%%%%%%%%%%%%%%%%
We perform the blocking transformation starting from points near the
attractor in the
$(\beta_{11},\beta_{12})$ plane,
i.e. the $\beta_{\rm twist}=0$ sector. The results are shown in
Fig. \ref{fig:3dim}.
%%%%%%%%%%%%%%%%%%%%%%%%%%%%%%%%%%%%%%%%%%%%%%%%%%%%
%%%%%%%%%%%%%%%%%%%%%%%%%%%%%%%%%%%%%%%%%%%%%%%%%%%%
As shown in the figure, no sizable value of $\beta_{\rm twist}$ is generated
for points on the attractor.
On the other hand, if we start from the points
$\beta_{11} \sim 10, \beta_{12} \sim -1.0 $ ,
a negative $\beta_{\rm twist}$ is induced. This
indicates that the attractor in three coupling space sits below the
$\beta_{{\rm twist}}=0$ sector in this region.
Although more extensive studies
are necessary to clarify the exact situation, it seems that
the attractor in two coupling space gives a good starting point for
designing improved actions.

\section{Summary}

We investigate the renormalization group (RG) flow of SU(3) lattice gauge
theory in a two coupling space, $(\beta_{11},\beta_{12})$.
An extensive numerical calculation of the RG flow on the lattice is made.
Swendsen's blocking followed by an effective action search using
the Schwinger-Dyson method is adopted to find renormalization effects.
%%%%%%%%%%%%%%%%%% Added  %%%%%%%%%%%%%%%%%%%%%%%%%%%%%%%%%%%%%%%%%%%%
In our previous analysis, we adopted the Demon method.
Although both methods give similar results, the Schwinger-Dyson algorithm
is more  echonomical because in case of the Demon formula we need
an additional Monte Carlo simulation for each configuration.
Moreover the Schwinger-Dyson method uses larger loops to determine the
couplings.
%%%%%%%%%%%%%%%%%%%%%%%%%%%%%%%%%%%%%%%%%%%%%%%%%%%%%%%%%%%%%%%%%%%%%%%

Analyses are performed in the fourth quadrant of the coupling space
and reveal the presence of an attractive stream.
Trajectories are first attracted towards this stream and afterwards
they move towards the origin along it.  The stream converges to a parabolic
curve
in the strong coupling region.  These features indicate that
the RT locates close to the two coupling space and the attractive stream
traces the RT.

A model flow equation which consists of asymptotic scaling, an irrelevant
coupling and a non-perturbative force corresponding to the area law
can reproduce the observed features.

%We also examine the scaling properties of actions near the attractive
%stream.
%In particular the one called  DBW2 (doubly blocked from Wilson action in
%2 coupling space) shows excellent restoration of rotational invariance.
%It should be noted, however, that the RT in the coupling regions
%analyzed here is not yet a single straight line.  Therefore the optimal
%$-\beta_{12}/\beta_{11}$ at weaker coupling is larger than for Symanzik
%or Iwasaki actions. The best improved
%actions are different at different coupling regions.

%%%%%%%%%    new %%%%%%%%%
In this paper we have compared several actions in the two coupling space
by measuring the restoration of the rotational symmetry and the scaling
of $\sqrt{\sigma}/T_c$.
In the regions of $a \sim 0.3$, Iwasaki and DBW2 actions, which are
near to RT , are superior to tree-level Symanzik and plaquette actions.
Although "improveness" of all actions are indistinguishable at smaller
lattice spacing within the present statistics, this indicate that
nonperturbative study of RG flows is valuable to design an improved action
at intermediate lattice spacing.
It is highly desirable to pursue the same analysis for fermion actions.
%%%%%%%%%%%%%%%%%%%%%%%%%%%

The effect of truncation to a space with only two couplings is partly
examined
and it is found
that renormalization effects outside the two coupling space are small
and the attractive stream in this space gives hence a good
starting  point for improvement.

\section{Acknowledgments}
All simulations have been done on CRAY J90
at Information Processing Center,
Hiroshima University,
SX-4 at RCNP, Osaka University
and on VPP500 at KEK (High Energy Accelerator Research Organization).
The authors would like to acknowledge M. Okawa for discussions on
the Schwinger-Dyson method. They acknowledge also G. Bali for
performing the measurement of the heavy quark potential.
H. Matsufuru would like to thank the Japan Society for the
Promotion of Science for financial support.
This work is supported by the Grant-in-Aide for
Scientific Research by Monbusho, Japan (No. 11694085)
and (No. 10640272).
This work is supported also by the Supercomputer Project No32 (FY1998)
of High Energy Accelerator Research Organization (KEK).

\end{document}